\documentclass{article}%

\usepackage{siunitx}
\usepackage{url}

\usepackage{amsmath}%
\usepackage{amsfonts}%
\usepackage{amssymb}%
\usepackage{graphicx}

\begin{document}

\title{Blueberry Earth}
\author{Anders Sandberg\\Oxford Martin School, University of Oxford}
\maketitle

\begin{abstract}
\noindent This paper explores the physics of the what-if question ``what if the entire Earth was instantaneously replaced with an equal volume of closely packed, but uncompressed blueberries?''. While the assumption may be absurd, the consequences can be explored rigorously using elementary physics. The result is not entirely dissimilar to a small ocean-world exoplanet.
\end{abstract}

\section{Introduction}

On the question-and-answer website Physics Stackexchange the user ``Billybodega'' asked the question\footnote{\url{https://physics.stackexchange.com/questions/418591/what-would-happen-if-the-earths-entire-volume-was-replaced-with-blueberries}}:

\begin{quotation}
\noindent  Supposing that the entire Earth was instantaneously replaced with an equal volume of closely packed, but uncompressed blueberries, what would happen from the perspective of a person on the surface?
\end{quotation}

Unfortunately the site tends to frown on fun what-if questions like this, so it was in my opinion prematurely closed while I was working out the answer. So here it is, with some extra extensions and corrections. 

The following modelling is more qualitative than quantitative, as befits a somewhat loose scenario. Still, it is possible to do a fairly rigorous analysis of what would occur and the properties of the resulting body.

\section{Initial state}

User WillO on Stackexchange estimated the density of blueberries to 13\% of Earth's density ($5510\times 0.13=\SI{716.3}{\kilo\gram\per\cubic\meter}$), while another estimate is~\SI{625.56}{\kilo\gram\per\meter\cubed},\footnote{\url{https://www.aqua-calc.com/page/density-table/substance/blueberries-coma-and-blank-raw}} and Rabbiteye blueberries lie in the range
\SIrange{782}{866}{\kilo\gram\per\meter\cubed} \cite{yemmireddy2013effect}. Assuming blueberry density to be around $ \rho_{\text{berries}}=\SI{700}{\kilo\gram\per\cubic\meter}$ appears to be reasonable. 

Note that these are  the big, thick-skinned highbush blueberries ({\em Vaccinium corymbosum}) rather than the small wild thin-skinned blueberries (bilberries, {\em V. myrtillus}) I grew up with; the latter would have higher unpacked density due to their smaller size and break far more easily.

Blueberry pulp has a density similar to water. One source gives~980 to~\SI{1050}{\kilo\gram\per\cubic\meter} \cite{mercali2011physical} although this is both temperature dependent and depends on how much solids there are -- water has been added. Skupie{\'n} mentions a water content of 80.1-87.7\% \cite{skupien2006chemical}, suggesting a density about 10\% different from water. We will assume $\rho_{\text{pulp}}=\SI{1000}{\kilo\gram\per\cubic\meter}$; while somewhat lower than the likely value we will only need a rough estimate.

The difference to the whole berries is due to the air between the berries. If the berries were hexagonally close packed spheres of water density the density would be about~\SI{740}{\kilo\gram\per\meter\cubed}, while the more likely random packing would be~\SI{634}{\kilo\gram\per\meter\cubed} \cite{song2008phase}. In the following we will ignore the mass of the air compared to the mass of the berries. 

So instantaneously turning Earth into blueberries will reduce its mass to~0.1274 of what it was. Gravity will become correspondingly weaker, $g_{\text{BE}}=0.1274 g=\SI{1.2510}{\meter\per\second\square}$.

\section{Coalescence}

However, blueberries are not particularly sturdy. While there is a literature on blueberry mechanics (of course!) \cite{donahue1998sensory,chiabrando2009mechanical,li2013miniature}, I did not manage to find a great source on their compressive strength. A rough estimate is possible: stacking a sugar cube (1 g) on a berry will not break it, while a milk carton (1 kg) will; 100 g has a decent but not certain chance. Indeed, this is not so far off from the slightly higher (\SIrange{178}{219}{\gram}) estimates in \cite{concha2015shelf}. So if we assume the blueberry area to be one square centimetre the breaking pressure is on the order of $ P_{\text{break}}=0.1 g / 10^{-4} \approx \SI{10000}{\newton\per\square\meter}$. This allows us to estimate at what depth the berries will start to break: it will happen at a depth $z$ where the pressure $\rho_{\text{berries}}g_{\text{BE}}$ from above reaches $P_{\text{break}}$. $$ z=\frac{P_{\text{break}}}{g_{\text{BE}}\rho_{\text{berries}}} = \SI{11.4198}{\meter}.$$ So while the surface will be free blueberries they will start pulping within a few meters below.

This pulping has an important effect: the pulp separates from the air, coalescing into a smaller sphere. If we assume pulp to be an incompressible fluid, then a sphere of pulp with the same mass as the initial berries will be $ \rho_{\text{pulp}} r_{\text{pulp}}^3 = \rho_{\text{berries}}r_{\text{earth}}^3$, or $$ r_{\text{pulp}} = \left(\frac{\rho_{\text{berries}}}{ \rho_{\text{pulp}} } \right)^{1/3}r_{\text{earth}}.$$ In this case we end up with a planet with 0.8879 times smaller radius (\SI{5663.2}{\kilo\meter}), surrounded by a vast atmosphere.

The freefall timescale for the planet (the time it would take for it to collapse under its own gravity if there were no other forces) is $$\tau = \sqrt{\frac{3\pi}{32G\rho_{\text{berries}}}},$$ 42 minutes. But relatively shortly the pulping interactions, air convection and other forces will slow things down in a complicated way. I expect that the the actual coalescence will take hours, with some late bubbles from the deep interior erupting fairly late.

\section{Atmosphere}

The gravity on the pulp surface is just $g_{\text{pulp}}=\SI{1.5832}{\meter\per\second\square}$, 16\% of normal gravity - almost exactly lunar gravity. This weakens convection currents and the speed with which bubbles move up since these involve forces proportional to $g$. The scale height of the atmosphere, $h=\frac{k_BT}{mg}$ where $m$ is the mean molecular mass, describes how pressure declines with height ($P(z)=P(0)e^{-z/h}$). Assuming the same composition and temperature as on Earth it will be 6.2 times higher, 53 km. This means that pressure will decline much less with altitude, allowing far thicker clouds and weather systems. As we will see, the atmosphere will puff up further due to heating.

The atmosphere is massive compared to Earth. The air content of the berries masses $M_{\text{air}}=\SI{1.0869e21}{\kilo\gram}$, 211 times the mass of Earth's atmosphere. The mass of a column of air will be $M_c=P(0)/g$, if we assume the atmosphere is still thin compared to the gravity well of the planet. Since we know that $M_{\text{air}} \approx 4\pi r_{\text{pulp}}^2 M_c$ we get $P(0)\approx M_{\text{air}}/4\pi r_{\text{pulp}}^2 g_{\text{pulp}}$, 16.8647 times Earth's air pressure. This kind of dense, deep atmosphere is somewhat similar to Titan's atmosphere. 

The colour of the atmosphere will be dominated by Rayleigh scattering, making it blueish, and the white from clouds. Since the optical depth is proportional to scale height, there will be several times more absorption and scattering than on Earth: the atmosphere is likely to be somewhat optically thick, making the ground light level only $\approx 4\%$ of Earth's.

\section{Heating}

The separation into pulp and air has big consequences. Enormous amounts of air will be pushing out from the pulp as bubbles and jets, producing spectacular geysers (especially since the gravity is low). Even more dramatic is the heating: a lot of gravitational energy is released as the mass is compacted. The total gravitational energy of a constant density sphere of radius $R$ is
$$\int_0^R \frac{G (4\pi r^2 \rho)(4 \pi r^3 \rho/3)}{r} dr  = \left(\frac{16\pi^2 G\rho^2}{3}\right ) \int_0^R r^4 dr $$
$$ =\left(\frac{16\pi^2 G}{15}\right)\rho^2 R^5.$$
The first factor in the integral is the mass of a spherical shell of radius r, the second the mass of the stuff inside, and the third the $1/r$ gravitational potential. 

If we ignore the mass of the air since it is small and we just want an order of magnitude estimate,  the compression of the berry mass gives energy
$$ E=\left(\frac{16\pi^2 G}{15}\right)\left(\rho_{\text{berries}}^2 r_{\text{earth}}^5 - \rho_{\text{pulp}}^2R_{\text{pulp}}^5\right ) \approx 4.5878\times \SI{e29}{\joule}.$$

This is the energy output of the sun over 20 minutes, nothing to sneeze at: blueberry earth will become hot. There is about~\SI{603}{\kilo\joule\per\kilo\gram}, more than enough to heat the blueberries from freezing to boiling. If blueberries have heat capacity equal to the one of water the temperature increase is 143.5793 degrees. When it reaches the boiling point things get more complicated, since there is not enough energy to boil off all the water (the latent heat of evaporation on Earth's surface is~\SI{2.257}{\mega\joule\per\kilo\gram}, able to boil off just a quarter of the water from already near boiling water). The added steam will also increase the pressure and hence make further boiling require higher temperature. 

The result is that blueberry earth will turn into a roaring ocean of boiling jam, with the geysers of released air and steam likely ejecting at least a few berries into orbit\footnote{Escape velocity is just~\SI{4.234}{\kilo\meter\per\second}, and berries at the initial surface will be even higher up in the potential. The air flows from deeper layers are subjected to significant pressure and heating, likely making them supersonic and hence able to loft material to close to escape velocity.}. As the planet evolves a thick atmosphere of released steam will add to the already considerable air from the berries. It is not inconceivable that the planet may heat up further due to a water vapour greenhouse effect, turning into a very odd Venusian world. There is also a chance that it reaches a state where water is supercritical, making the ocean fade into the atmosphere. This would require~\SI{647.096}{\kelvin} and 217.755 Earth-atmospheres of pressure; this appears possible since it only requires 2\% of the water to form steam.

\section{The ice core}

Meanwhile the jam ocean is very deep, and the pressure at depth will be enough to cause the formation of high pressure ice even if it is warm. If the formation process is slow there will be some separation of water into ice and a concentration of other chemicals in the jam ocean, but I suspect the rapid collapse will instead make some kind of composite pulp ice. Ice VII forms above 3 GPa, so if we just naively use constant gravity this happens at a depth $$ z_{ice}=\frac{P_{\text{VII}}}{g_{\text{pulp}}\rho_{\text{pulp}}}\approx \SI{1895}{\kilo\meter},$$ about two-thirds of the radius. This would make up most of the interior. 

However, this is misleading. Gravity is a bit weaker in the interior since there is less mass below a certain radius than at the surface, so we need to take that into account. The pressure due to all the matter above radius $r$ is $$ P(r) =\left(\frac{3GM^2}{8\pi R^4}\right)\left(1-\left(\frac{r}{R}\right)^2\right),$$ and the ice core will hence have radius $ r_{ice}=\sqrt{1-\frac{P_{\text{VII}}}{P(0)}}R \approx \SI{3257}{\kilo\meter}$. This is smaller, about 57\% of the radius, and just 20\% of the total volume.

\section{Other effects}

The coalescence will also speed up rotation. The original blueberry earth would of course make one rotation every 24 hours, but the smaller result would have a smaller moment of inertia. Angular momentum conservation gives $(2/5)MR_1^2(2\pi/T_1) = (2/5)MR_2^2(2\pi/T_2)$, or $$ T_2 = \left(\frac{R_2}{R_1}\right)^2 T_1,$$ in this case~18.9210 hours. This in turn will increase the oblateness a bit, to approximately\footnote{\url{http://farside.ph.utexas.edu/teaching/336k/Newton/node109.html}} 0.038 -- an 8.8 times increase over Earth, but still not even comparable to Jupiter (0.06487). Due to angular momentum conservation the axial tilt will not change. 

The magnetic field would disappear: there is no longer any iron core to power it. This in turn would remove the magnetosphere, making high-energy solar and cosmic particles free to hit the outer atmosphere. This will in turn lead to the dissociation of molecules and in particularly the escape of hydrogen from any water vapour up there. This process is likely to lead to the gradual drying of the planet as the water disappears unless there is a cold trap in the stratosphere to hold it down, although it might be slow enough to not matter over the remaining main sequence lifespan of the sun \cite{kuchner2003volatile}.

Another effect is the orbit of the Moon. Now the two bodies have about equal mass. Is the Moon bound to blueberry earth? A kilogram of lunar material has potential energy $GM_{\text{BE}}/r_{\text{moon}} \approx \SI{1.3209e5}{\joule}$, while the kinetic energy is~\SI{5.2341e5}{\joule} -- more than enough to escape. Had it remained the jam ocean would have made an excellent tidal dissipation mechanism that would have slowed down rotation and moved blueberry earth towards tidal lock with the moon much earlier than the 50 billion years it would otherwise have taken \cite[p. 184]{murray1999solar}.

\section{Summary}

So, to sum up, to a person standing on the surface of the Earth when it turns into blueberries, the first effect would be a drastic reduction of gravity. Standing on the blueberries might be possible in theory, except that almost immediately they begin to compress rapidly and air starts erupting everywhere. The effect is basically the worst earthquake ever, and it keeps on going until everything has fallen 715 km. While this is going on everything heats up drastically until the entire environment is boiling jam and steam. The end result is a world that has a steam atmosphere covering an ocean of jam on top of warm blueberry granita.

The final state of Blueberry Earth is somewhat similar to oceanic exoplanets \cite{leger2004new,kuchner2003volatile}, although far lighter than any observed so far.

Assuming a somewhat denser pulp density does not radically change the conclusions; the result is somewhat smaller and hotter. Reinterpreting the question as starting with densely packed blueberries removes the main energy release from coalescence and the addition of air, but there will likely be a sedimentation process releasing a fraction of the energy. In this case Earth just turns into a waterworld and retains the Moon.

There are still many things to explore. What is the likely chemistry of the blueberry world? On one hand blueberries are full of antioxidants ($\approx 0.2\%$ by weight) \cite{skupien2006chemical}, but on the other there is also a significant amount of oxidable compounds such as sugars (10\%) and an extensive and hot atmosphere. Most likely there will be massive oxidation process, perhaps fermenting the sugars into alcohols. Will the different components of the pulp separate and form layers around the ice core? What about simulating the internal structure with more detailed equations of state? Will the moist and deep atmosphere heat up further, or could cloud layer keep the albedo high enough to avoid a runaway greenhouse? Will there remain a cold trap in the stratosphere preventing water from being dissociated in the ionosphere, or will blueberry earth gradually dry out? Could extremophile bacteria survive and bootstrap an ecosystem?

One might wonder if this kind of exploration is worthwhile. I believe it is: this is both a pedagogical and amusing way of applying standard planetary science modelling to a system. Given how exotic exoplanets have turned out, the physics of blueberry earth is actually fairly normal compared to much that is out there.

\bibliography{blueberryearth} 

\begin{thebibliography}{10}

\bibitem{chiabrando2009mechanical}
V.~Chiabrando, G.~Giacalone, and L.~Rolle.
\newblock Mechanical behaviour and quality traits of highbush blueberry during
  postharvest storage.
\newblock {\em Journal of the Science of Food and Agriculture}, 89(6):989--992,
  2009.

\bibitem{concha2015shelf}
A.~Concha-Meyer, J.~D. Eifert, R.~C. Williams, J.~E. Marcy, and G.~E. Welbaum.
\newblock Shelf life determination of fresh blueberries (vaccinium corymbosum)
  stored under controlled atmosphere and ozone.
\newblock {\em International journal of food science}, 2015, 2015.

\bibitem{donahue1998sensory}
D.~W. Donahue and T.~M. Work.
\newblock Sensory and textural evaluation of maine wild blueberries for the
  fresh pack market.
\newblock {\em Journal of texture studies}, 29(3):305--312, 1998.

\bibitem{kuchner2003volatile}
M.~J. Kuchner.
\newblock Volatile-rich earth-mass planets in the habitable zone.
\newblock {\em The Astrophysical Journal Letters}, 596(1):L105, 2003.

\bibitem{leger2004new}
A.~L{\'e}ger, F.~Selsis, C.~Sotin, T.~Guillot, D.~Despois, D.~Mawet,
  M.~Ollivier, A.~Lab{\`e}que, C.~Valette, F.~Brachet, et~al.
\newblock A new family of planets? "ocean-planets".
\newblock {\em Icarus}, 169(2):499--504, 2004.

\bibitem{li2013miniature}
C.~Li, P.~Yu, F.~Takeda, and G.~Krewer.
\newblock A miniature instrumented sphere to understand impacts created by
  mechanical blueberry harvesters.
\newblock {\em HortTechnology}, 23(4):425--429, 2013.

\bibitem{mercali2011physical}
G.~D. Mercali, J.~R. Sarkis, D.~P. Jaeschke, I.~C. Tessaro, and L.~D.~F.
  Marczak.
\newblock Physical properties of acerola and blueberry pulps.
\newblock {\em Journal of Food Engineering}, 106(4):283--289, 2011.

\bibitem{murray1999solar}
C.~D. Murray and S.~F. Dermott.
\newblock {\em Solar system dynamics}.
\newblock Cambridge university press, 1999.

\bibitem{skupien2006chemical}
K.~Skupie{\'n}.
\newblock Chemical composition of selected cultivars of highbush blueberry
  fruit (vaccinium corymbosum l.).
\newblock {\em Folia Horticulturae}, 18(2):47--56, 2006.

\bibitem{song2008phase}
C.~Song, P.~Wang, and H.~A. Makse.
\newblock A phase diagram for jammed matter.
\newblock {\em Nature}, 453(7195):629, 2008.

\bibitem{yemmireddy2013effect}
V.~K. Yemmireddy, M.~S. Chinnan, W.~L. Kerr, and Y.-C. Hung.
\newblock Effect of drying method on drying time and physico-chemical
  properties of dried rabbiteye blueberries.
\newblock {\em LWT-Food Science and Technology}, 50(2):739--745, 2013.

\end{thebibliography}
\bibliographystyle{abbrv}

\end{document}